\documentclass[aps,prb,reprint,groupedaddress]{revtex4-2}

\usepackage{dcolumn}
\usepackage{bm}
\usepackage[section]{placeins}
\usepackage{color}
\usepackage{mathrsfs}
\usepackage{ulem}
\usepackage{hyperref}
\usepackage{graphicx}
\usepackage{svg}
\usepackage{braket}
\usepackage{here}
\usepackage{placeins}
\usepackage{float}
\usepackage{siunitx}

\begin{document}


\title{Diamond quantum-sensing platform with integrated boron-doped diamond microwave antenna and thermometer}

\author{Masahiro Ohkuma$^{1}$}
\email{okuma.m.408d@m.isct.ac.jp}
\author{Ryo Matsumoto$^{2}$}
\author{Shintaro Adachi$^{3}$}
\author{Shinobu Onoda$^{4}$}
\author{Kensei Terashima$^{2}$}
\author{Takao Watanabe$^{5}$}
\author{Yoshihiko Takano$^{2}$}
\author{Keigo Arai$^{1}$}
\affiliation{$^{1}$School of Engineering, Institute of Science Tokyo, Yokohama 226-8501, Kanagawa, Japan}
\affiliation{$^{2}$Research Center for Materials Nanoarchitectonics (MANA), National Institute for Materials Science, Tsukuba 305-0047, Ibaraki, Japan}
\affiliation{$^{3}$Department of Mechanical and Electrical Systems Engineering, Kyoto University of Advanced Science (KUAS), Kyoto 615-8577, Kyoto, Japan}
\affiliation{$^{4}$National Institutes for Quantum Science and Technology (QST), Takasaki 370-1292, Gunma, Japan}
\affiliation{$^{5}$Graduate School of Science and Technology, Hirosaki University, Hirosaki 036-8561, Aomori, Japan}
\date{\today}

\begin{abstract}
Wide-field nitrogen-vacancy (NV) magnetic imaging at cryogenic temperatures requires microwave excitation and reliable knowledge of the temperature near the sensing region.
Here, we report an integrated diamond quantum-sensing platform combining an ensemble of NV centers with a boron-doped diamond (BDD) microwave antenna and thermometer formed on the same diamond substrate.
The BDD antenna provides microwave excitation for optically detected magnetic resonance measurements, and the BDD thermometer monitors the thermal environment near the NV sensing region.
The BDD thermometer detected laser-induced local heating that was not clearly resolved by a stage-mounted thermometer.
Using this platform, we imaged the temperature-dependent Meissner response of multiple cuprate superconductors while recording the temperature.
These results demonstrate that the integrated BDD--NV platform provides a practical approach for cryogenic wide-field magnetic imaging with integrated microwave delivery and local thermometry.
\end{abstract}

\maketitle

Nitrogen-vacancy (NV) centers in diamonds have emerged as a platform for magnetometry in condensed-matter physics \cite{taylorHighsensitivityDiamondMagnetometer2008a, degenQuantumSensing2017a, casolaProbingCondensedMatter2018}.
They have been used to visualize local magnetic fields in superconductors, magnetic materials, and current-carrying devices \cite{thielQuantitativeNanoscaleVortex2016a,nishimuraWidefieldQuantitativeMagnetic2023, dovzhenkoMagnetostaticTwistsRoomtemperature2018, thielProbingMagnetism2D2019, kuImagingViscousFlow2020, garsiThreedimensionalImagingIntegratedcircuit2024a}.
Owing to their optically addressable spin states, NV centers enable noninvasive magnetic imaging with a high spatial resolution \cite{gruberScanningConfocalOptical1997,dohertyNitrogenvacancyColourCentre2013}.
Furthermore, their spin coherence can be maintained under extreme conditions, including high temperatures \cite{toyliMeasurementControlSingle2012, liuCoherentQuantumControl2019, fanQuantumCoherenceControl2024} and high pressure \cite{bhattacharyyaImagingMeissnerEffect2024, wangImagingMagneticTransition2024}, making them attractive for quantum sensing in a wide range of environments.
Extending such environmental durability to the microwave-delivery structure is also important for quantum sensing under extreme conditions. Recently, we demonstrated the coherent control of NV spins using a boron-doped diamond (BDD) microwave antenna and verified its operation under high-temperature, cryogenic, and high-pressure conditions, indicating that BDD can serve as a robust microwave-delivery element for NV-based quantum sensing \cite{ohkumaCoherentControlSolidstate2026,ohkumaWidefieldNVMagnetometry2026}.

For low-temperature NV-based quantum sensing, accurate knowledge of the local temperature near the sensing region is crucial.
Laser illumination for spin initialization and readout, as well as microwave excitation for spin manipulation, can induce local heating in both the diamond sensor and sample.
Although NV centers can be used for thermometry through the temperature dependence of their zero-field splitting, the temperature sensitivity decreases substantially at cryogenic temperatures \cite{acostaTemperatureDependenceNitrogenVacancy2010, kucskoNanometrescaleThermometryLiving2013, chenTemperatureDependentEnergy2011, dohertyTemperatureShiftsResonances2014, cambriaPhysicallyMotivatedAnalytical2023}.
Therefore, an independent thermometer located in close proximity to the NV-sensing region is highly desirable for quantitative low-temperature measurements.

\begin{figure}[htb!]
\centering\includegraphics[]{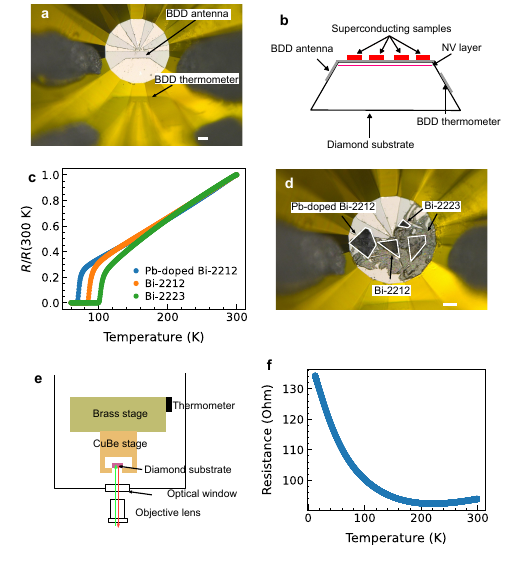}
\caption{\label{f1} (a) Optical image of the diamond substrate with boron-doped diamond antenna and thermometer.
(b) Schematic of the diamond substrate with a boron-doped diamond antenna, thermometer, and superconducting samples.
(c) Temperature dependence of the resistance of Bi-based cuprate superconductors.
The resistance values were normalized to their respective values at 300 K.
(d) Optical image of the superconducting samples on the diamond substrate.
(e) Schematic of the low-temperature setup for wide-field ODMR measurements.
(f) Temperature dependence of the resistance of the boron-doped diamond thermometer.
The scale bars in (a) and (d) represent 50 $\mu$m}
\end{figure}

In this manuscript, we report an integrated diamond quantum-sensing platform that combines an ensemble of NV centers with BDD structures for microwave delivery and thermometry.
The BDD antenna delivers microwave excitation for wide-field optically detected magnetic resonance (ODMR) measurements, and the BDD thermometer monitors the temperature near the NV sensing region.
Using this integrated NV--BDD platform, we performed wide-field magnetic imaging of multiple superconducting samples while monitoring the local temperature.
The demonstrated design provides a practical approach for cryogenic wide-field magnetic imaging with integrated microwave delivery and local thermometry.

We used a type Ib diamond substrate with a (100)-oriented surface (SYNTEK Co., Ltd.) for the experiment.
Vacancies were introduced into the diamond substrate by ${}^{12}{\rm C}^{+}$ ion implantation at an energy of 30~keV and fluence of $5\times10^{12}~{\rm {cm}}^{-2}$.
Subsequently, the implanted substrate was annealed in vacuum at $\SI{1000}{\degreeCelsius}$ for 2~h to form NV centers.
Simulations using the Stopping and Range of Ions in Matter software tools suggest that the vacancies created by implantation are spread within a distance of approximately 50 nm from the surface of the diamond \cite{zieglerSRIMStoppingRange2010}.
After the formation of NV centers, the BDD antenna and thermometer were grown by microwave plasma chemical vapor deposition \cite{takanoSuperconductivityDiamondThin2004, matsumotoNoteNovelDiamond2016}.
An optical image and schematic of the diamond substrate are shown in Fig.~\ref{f1}a and b, respectively.

Single crystals of Bi-based cuprate superconductors, including Bi$_2$Sr$_2$CaCu$_2$O$_{8+\delta}$ (Bi-2212), Pb-doped Bi-2212, and Bi$_2$Sr$_2$Ca$_2$Cu$_3$O$_{10+\delta}$ (Bi-2223), were grown by the traveling-solvent floating-zone method \cite{fujiiSinglecrystalGrowthBi2Sr2Ca2Cu3O10+d2001, fujiiDopingDependenceAnisotropic2002, adachiSinglecrystalGrowthUnderdoped2015, adachiUnscalingSuperconductingParameters2015}.
To prepare three samples with distinctly different superconducting transition temperatures ($T_{\rm c}$), oxygen annealing was utilized to tune the doping states: Pb-doped Bi-2212 was brought into an overdoped state ($T_{\rm c}\sim68$ K), Bi-2212 into a slightly overdoped state ($T_{\rm c}\sim84$ K), and Bi-2223 into a nearly optimally doped state ($T_{\rm c}\sim98$ K) \cite{adachiUnscalingSuperconductingParameters2015}.
Figure~\ref{f1}c shows the temperature dependence of the resistance of representative crystals prepared under the same growth and annealing conditions.
Each $T_{\rm c}$ was defined as the temperature at which zero electrical resistance was observed in conventional DC four-terminal measurements using the Physical Property Measurement System (Quantum Design).
An optical image of the samples on the diamond substrate is shown in Fig.~\ref{f1}d.

Wide-field ODMR measurements under cryogenic conditions were carried out with a lab-built optical setup combined with a custom-built Gifford--McMahon refrigerator.
The temperature dependence of the resistance of the BDD thermometer was measured using a four-terminal method with a current source (6146, ADCMT) and voltmeter (34401A, Agilent).
The temperature was measured using a thermometer attached to a brass stage, to which a CuBe stage supporting the diamond substrate was connected.
A schematic of the sample stage is presented in Fig.~\ref{f1}e.
The NV centers were excited using a 532-nm laser (MLL-S-532B, CNI Laser) focused through an objective lens (M-PLAN APO 7.5X, Mitutoyo). 
The emitted red fluorescence was collected through the same objective, filtered using a long-pass filter, and imaged on an electron multiplying charge-coupled device camera (iXon Ultra 897, Andor). 
Microwaves from a signal generator (SynthHD, Windfreak) were amplified by a power amplifier (ZHL-16W-43-S$+$, Mini-Circuits) and controlled using a microwave switch (ZASWA-2-50DR$+$, Mini-Circuits) triggered by a data acquisition device (NI 6363, National Instruments).
During wide-field ODMR measurements, fluorescence images were acquired sequentially under microwave-off and -on conditions.
To minimize mechanical vibrations, the cold head was temporarily turned off during wide-field ODMR measurements, which caused the temperature to gradually increase during acquisition.
A magnetic field of approximately 2.0 mT was applied along the $\langle100\rangle$ direction using a coil wound around the CuBe stage.

\begin{figure}[htb!]
\centering\includegraphics[]{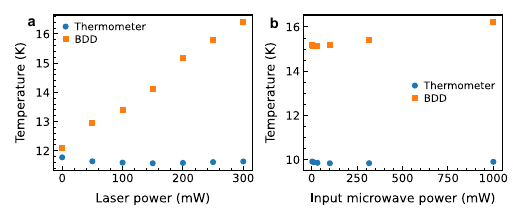}
\caption{\label{f2} Local thermometry using the integrated BDD thermometer.
(a) Laser-power dependence of the temperature measured by the stage-mounted thermometer and BDD thermometer.
(b) Microwave-power dependence of the temperature under continuous laser irradiation at 300 mW.
The microwave frequency was set to 2.87 GHz.
The horizontal axis represents the nominal input microwave power estimated from the microwave-source output setting and amplifier gain.}
\end{figure}

Figure~\ref{f1}f shows the temperature dependence of the electrical resistance of the BDD thermometer. The resistance exhibited a non-monotonic temperature dependence, with a minimum at approximately 220 K.
To determine the temperature from the BDD resistance in the range relevant to the present measurements, the data between 10 and 120 K were fitted with a fourth-order polynomial.
The resulting fitting curve was used to convert the measured resistance into temperature.

Figure~\ref{f2}a compares the temperatures measured by the stage-mounted thermometer and integrated BDD thermometer as a function of the laser power.
Although the stage temperature remained nearly constant, the BDD thermometer detected a temperature increase exceeding 4 K at the highest laser power.
This result indicates that laser-induced local heating occurred near the diamond sensor and was not adequately captured by the stage thermometer.
The small offset between the two thermometers before laser irradiation was attributed to uncertainties in the resistance-to-temperature calibration of the BDD thermometer.

\begin{figure*}[htb!]
\centering\includegraphics[]{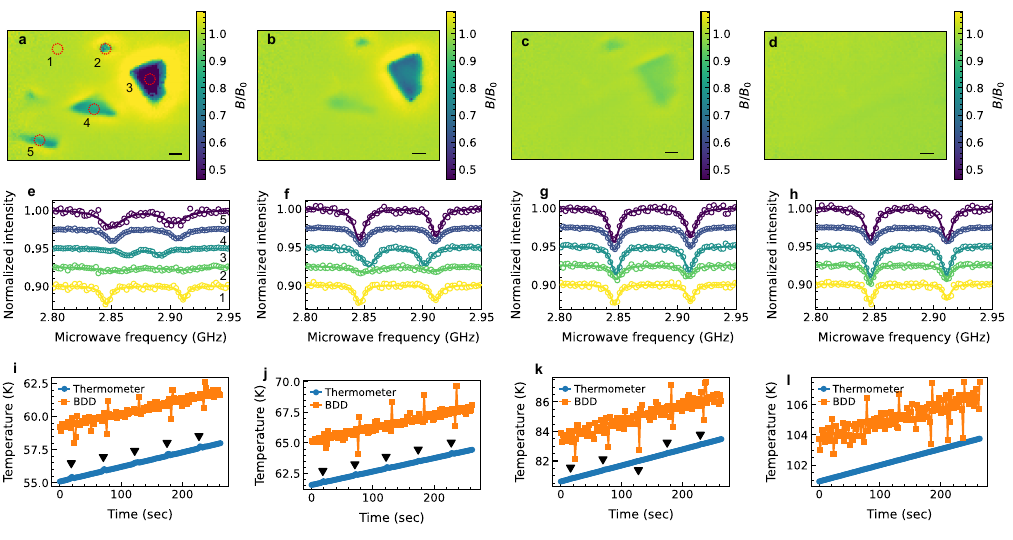}
\caption{\label{f3} Temperature-dependent wide-field ODMR imaging of superconducting samples with simultaneous local thermometry.
(a)--(d) Magnetic-field maps normalized by the applied magnetic field acquired at different temperatures.
Four superconducting samples are included in the field of view; the diamagnetic signals disappeared sequentially as the temperature increased.
Scale bars indicate 20 $\mu$m.
(e)--(h) Representative ODMR spectra measured at the positions indicated by the red dotted circles and labels 1--5 in the corresponding magnetic-field maps.
The ODMR spectra are vertically offset by 0.025 for clarity.
(i)--(l) Temperature traces recorded during the corresponding ODMR measurements.
The local temperature measured by the integrated BDD thermometer is compared with the stage temperature measured by the stage-mounted thermometer.
The black triangles indicate transient temperature increases observed by the stage-mounted thermometer during the ODMR acquisition.}
\end{figure*}

In addition, we measured the microwave-power dependence of the temperature at 2.87 GHz under continuous laser irradiation at 300 mW, as shown in Fig.~\ref{f2}b.
Here, the nominal input microwave power was estimated by adding the amplifier gain to the output power setting of the microwave source.
The microwave power was not measured using a spectrum analyzer or power meter.
Moreover, owing to reflection, impedance mismatch, and transmission losses, the actual microwave power delivered to the BDD antenna was expected to be smaller than the nominal value.
Similar to the laser-power dependence, the stage-mounted thermometer showed almost no temperature change, whereas the BDD thermometer exhibited a clear temperature increase at higher nominal microwave powers.

Figure~\ref{f3}a--d shows the normalized magnetic-field maps acquired at different temperatures. Figure~\ref{f3}e--h shows the ODMR spectra measured at selected positions in the corresponding maps, and Fig.~\ref{f3}i--l shows the temperature traces recorded during each measurement.
All wide-field ODMR measurements were performed with a 300-mW laser power and 100-mW nominal microwave input power.
Each magnetic-field map contained four superconducting samples.
At low temperatures, all samples exhibited diamagnetic stray-field signals associated with the Meissner effect.
As the temperature increased, these signals disappeared successively, beginning with samples having lower superconducting transition temperatures.

During the ODMR measurements, the stage-mounted thermometer occasionally exhibited transient increases in temperature.
Moreover, the BDD thermometer signal showed apparent fluctuations of approximately $\pm 2$ K, including spike-like features.
These observations demonstrate that the integrated thermometer can continuously monitor the local thermal environment during cryogenic NV imaging.
The simultaneous acquisition of magnetic images and local temperature information is expected to be useful for the quantitative characterization of temperature-dependent phenomena in superconductors and other quantum materials.
These observations highlight the importance of local thermometry in cryogenic NV measurements.

As shown in Fig.~\ref{f2}b, no clear temperature increase was observed with either the BDD thermometer or stage-mounted thermometer at a nominal 100-mW input microwave power when the microwave frequency was fixed at 2.87 GHz.
By contrast, transient temperature increases were observed during wide-field ODMR measurements, although these measurements were performed at temperatures above 55 K using the same nominal microwave power.
This difference suggests that the observed heating depends on the microwave frequency.
Frequency-dependent transmission, reflection, and impedance matching in the microwave circuit can alter the actual microwave power delivered to the BDD antenna during the ODMR sweep, potentially resulting in enhanced heating at particular frequencies.
The spike-like variations in the BDD signal may also include contributions from microwave-induced electrical perturbations and may therefore not solely represent actual temperature changes.

\begin{figure}[htb!]
\centering\includegraphics[]{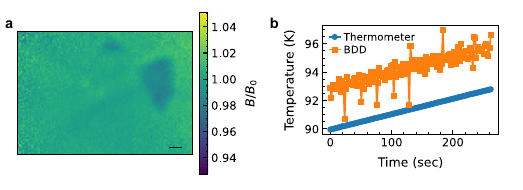}
\caption{\label{f4} (a) Magnetic-field maps normalized by the applied magnetic field, acquired at a temperature of $94.3\pm1.1$ K as measured by the BDD thermometer. The scale bar indicates 20 $\mu$m.
(b) Temperature traces recorded during the corresponding ODMR measurements.}
\end{figure}

Figure~\ref{f4} shows magnetic-field maps acquired at $94.3\pm1.1$ K as measured by the BDD thermometer.
The diamagnetic response of Bi-2223 persisted up to this temperature, consistent with our previous wide-field NV magnetic-imaging result obtained using a Pt-foil microwave antenna \cite{ohkumaProbingMeissnerEffect2026}.
This agreement suggests that microwave-induced heating due to the BDD antenna did not significantly affect the transition temperature determined in the present measurements.

The integrated BDD thermometer provides a practical means of detecting local heating induced by laser and microwave excitation for NV-based quantum sensing.
In principle, a separate thermometer can be attached directly to the diamond substrate; however, the BDD approach offers several advantages.
Because BDD can be patterned by microfabrication, the thermometer can be placed in close proximity to the NV sensing region.
In addition, because the BDD layer is grown directly on the diamond substrate, close thermal coupling to the diamond is expected, enabling the sensitive detection of local temperature changes near the NV centers.
These features make the BDD thermometer particularly useful for determining the temperature during low-temperature wide-field NV imaging.

A limitation of the present BDD thermometer is that its resistance shows a minimum at approximately 220 K, which prevents the absolute temperature from being uniquely determined between 10 and 300 K.
However, the electrical resistance can be controlled by the boron concentration to obtain a monotonic temperature dependence over the relevant temperature range \cite{takanoSuperconductivityCVDDiamond2009}.
Such optimization may also extend the applicable temperature range, including high-temperature operation \cite{matsumotoDiamondAnvilCell2021, matsumotoEmergenceSuperconductivity202025}.
Additionally, improving the temporal resolution of the BDD readout would enable more accurate tracking of temperature fluctuations during the NV measurements.
A faster readout would be useful for resolving transient temperature changes induced by laser and microwave excitations.
Such improvements would be particularly valuable in cryogenic studies of superconductors and other quantum materials, where small temperature variations can significantly influence the observed magnetic responses.

Recent advances in automated and autonomous experimentation have accelerated the discovery and optimization of functional materials \cite{steinProgressProspectsAccelerating2019, stachAutonomousExperimentationSystems2021, tomSelfDrivingLaboratoriesChemistry2024, nishioDigitalLaboratoryModular2025, wangOrchestrationHeterogeneousExperimental2026}.
In such closed-loop workflows, characterization tools are required to provide reproducible, scalable, and machine-readable feedback in addition to automated synthesis and processing.
Magnetic properties are one of the fundamental characteristics that should be addressed in the pursuit of functional materials such as superconductors. In this regard, the present wide-field NV platform is promising.
Wide-field NV magnetic imaging can be particularly powerful because it captures local magnetic fields over a wide field of view in parallel rather than by a point-by-point scan.
This advantage enables the simultaneous measurement of multiple samples under a common measurement configuration and thermal conditions.
In addition, this imaging technique can be used for spatially resolved measurements of device structures.
The robust BDD antenna provides microwave delivery under a wide range of measurement environments, whereas the integrated BDD thermometer provides temperature information for interpreting temperature-dependent magnetic responses.
Overall, these features suggest the potential of the platform for spatially resolved, temperature-dependent, and high-throughput magnetic property measurements.

In conclusion, we developed an integrated diamond quantum-sensing platform that combines an ensemble of NV centers with a BDD microwave antenna and thermometer. 
The BDD thermometer, which was grown directly on the diamond substrate, enabled temperature monitoring near the NV sensing region. 
By comparing the BDD thermometer with a stage-mounted Cernox thermometer, we demonstrated that the integrated BDD thermometer can detect temperature changes induced by laser and microwave excitation that are not clearly detected by the stage-mounted thermometer.
Using this platform, we performed wide-field ODMR imaging of multiple superconducting samples and observed the temperature-dependent disappearance of Meissner-induced magnetic signals while monitoring the temperature. 
These results show that integrated thermometry provides local temperature information for interpreting temperature-dependent material responses in low-temperature NV-based quantum sensing.
In addition, the present approach may be useful for future automated magnetic characterizations in autonomous materials exploration workflows.

\section*{DATA AVAILABILITY}
The data that support the findings of this study are available from the corresponding author upon reasonable request.

\begin{acknowledgments}
This work is supported by, or in part by, JSPS KAKENHI Grant numbers JP26H02233, JP25K01508, JP25K17937, JP24KJ1035, JP23KK0267, JP23K26528.
This work was also supported by JST ASPIRE Grant number JPMJAP24C1 and JACI Prize for Encouraging Young Researcher.
\end{acknowledgments}


\bibliography{NV}

\end{document}